# THE OPTICAL GRAVITATIONAL LENSING EXPERIMENT. ECLIPSING BINARIES AND SX PHE STARS IN $\omega$ CEN AND 47 TUC

J. KALUZNY, M. KUBIAK, M. SZYMAŃSKI, A. UDALSKI [1]
Warsaw University Observatory, Al. Ujazdowskie 4, 00-478 Warsaw

W. KRZEMIŃSKI
Las Campanas Observatory, Casilla 601, La Serena, Chile

MARIO MATEO [1]
Department of Astronomy, 821 Dennison Building, Univ. of Michigan, Ann Arbor, MI 48109

ABSTRACT    Extensive monitoring of the central part of $\omega$ Cen has lead to the discovery of 12 eclipsing binaries and 9 SX Phe stars. We report also finding 6 contact binaries in the field of 47 Tuc.

The OGLE experiment is a long-term project with the main goal of searching for dark matter in our Galaxy by identifying microlensing events toward the galactic bulge (Udalski *et al.* 1992). At times when the bulge is unobservable we perform other photometric programs (eg. Paczyński *et al.* 1995). The OGLE project is conducted using the 1-m Swope telescope at Las Campanas Observatory. A single 2K CCD giving a field of view of $14.5 \times 14.5$ $(')^2$ is used as a detector. During 1993 and 1994 seasons we monitored the globular clusters NGC 104 (=47 Tuc) and NGC 5139 (=$\omega$ Cen) in a search for variable stars. Of primary interest was the detection of detached eclipsing binaries which can potentially provide vital information about masses of stars in globular clusters. The data were collected in 3 fields covering the central part of $\omega$ Cen and in 3 fields located close to the center of 47 Tuc. Most observations were collected in the V-band. The number of frames obtained for a given field ranged from about 250 to about 500. 22 variables were identified in $\omega$ Cen: 9 SX Phe stars; 7 EW systems; 5 EA or EB systems. In Fig. 1 we show the location of $\omega$ Cen variables on the cluster CMD. Most variables are located among the blue stragglers. Eclipsing systems V15 and V17 are located in the transition region between main-sequence stars and subgiants. The light curves of these binaries (Fig. 2) indicate that they are well detached systems, with periods of 1.497 days and 2.467 days for V15 and V17, respectively. Having in hand accurate radial velocities and light curves of these binaries, one can hope to determine directly the masses of turnoff stars in $\omega$ Cen. We discovered 6 contact binaries in the field of 47 Tuc – all of them are blue stragglers.

---

[1] Visiting Astronomer, Las Campanas Observatory

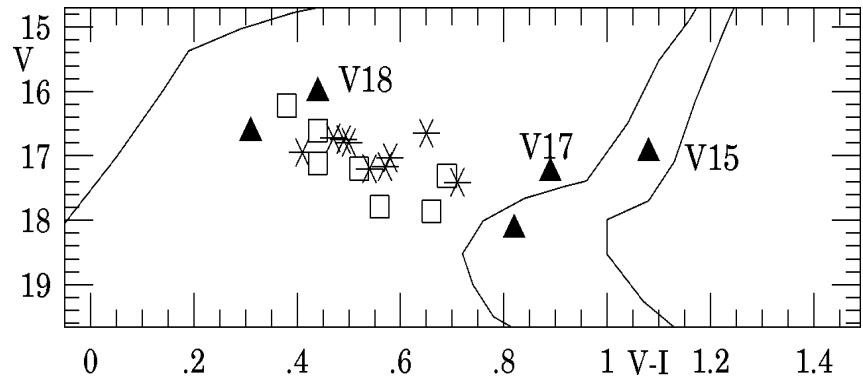

FIGURE I   A schematic CMD of $\omega$ Cen with marked positions of variables discovered by the OGLE group. Triangles – EA systems; squares – contact binaries ; asterisks – SX Phe stars. Variable V18 has already been discovered by Niss *et al.* (1978).

**REFERENCES**

Niss, B., Jorgensen, H.E., & Lautsen, S. 1978, A&AS, 32, 387

Paczyński, B. et al. 1995, IUA Symp. 169, "Unsolved Problems of the Milky Way", ed. L. Blitz, in print

Udalski, A., Szymański, M., Kaluzny, J., Kubiak, M., & Mateo, M. 1992, Acta Astron., 42, 253

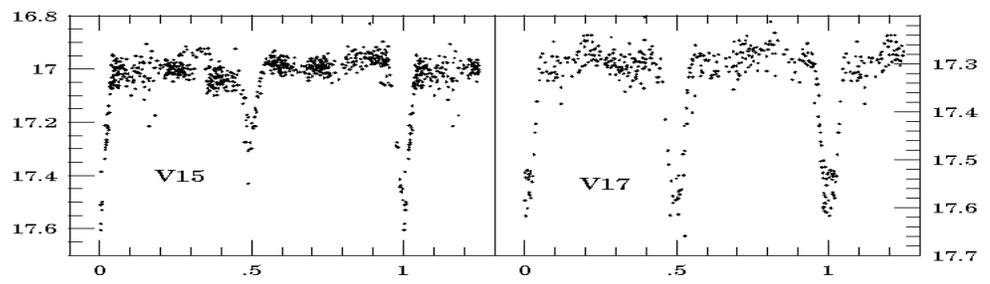

FIGURE II    Phased V-band light curves of $\omega$ Cen variables V15 and V17.